# A simple atmospheric electrical instrument for educational use

A.J. Bennett[1] and R.G. Harrison

*Department of Meteorology, The University of Reading*
*P.O. Box 243, Earley Gate, Reading RG6 6BB, UK*

**Abstract**
Electricity in the atmosphere provides an ideal topic for educational outreach in environmental science. To support this objective, a simple instrument to measure real atmospheric electrical parameters has been developed and its performance evaluated. This project compliments educational activities undertaken by the Coupling of Atmospheric Layers (CAL) European research collaboration. The new instrument is inexpensive to construct and simple to operate, readily allowing it to be used in schools as well as at the undergraduate University level. It is suited to students at a variety of different educational levels, as the results can be analysed with different levels of sophistication. Students can make measurements of the fair weather electric field and current density, thereby gaining an understanding of the electrical nature of the atmosphere. This work was stimulated by the centenary of the 1906 paper in which C.T.R. Wilson described a new apparatus to measure the electric field and conduction current density. Measurements using instruments based on the same principles continued regularly in the UK until 1979. The instrument proposed is based on the same physical principles as C.T.R. Wilson's 1906 instrument.

**Keywords**: electrostatics; potential gradient; air-earth current density; meteorology;

[1] *E-mail: a.j.bennett@reading.ac.uk*

## 1. Introduction

The phenomena of atmospheric electricity provide an ideal topic for stimulating lectures, talks and laboratory demonstrations. Educational measurements in the real atmosphere are, however, more troublesome to arrange, as the life-threatening hazards associated with lightning generally prevent their use in quantitative demonstrations. The safer fair weather circumstances require highly sensitive apparatus to detect the weak electric fields always present in such conditions. In the early days of atmospheric electricity, the mechanical electrometer and flame probe sensor provided a simple and direct method for illustrating the existence of the fair weather electric field. A ready example of this was provided by Lord Kelvin (Thomson, 1859)

> *"The author gave the result of a determination which he had made, with the assistance of Mr Joule, on the Links, a piece of level ground near the sea, beside the city of Aberdeen, about 8am on the preceding day (September 14), under a cloudless sky, and with a light north-west wind blowing, with the insulating stand of the collecting part of the apparatus buried in the ground, and the electrometer removed to a distance of 5 or 6 yards and connected by a fine wire with the collecting conductor. The height of the match was 3 feet above the ground, and the observer at the electrometer lay on the ground to render the electrical influence of his own body on the match insensible. The result showed a difference of potentials between the earth (negative) and the air (positive) at the match equal to that of 115 elements of Daniel's battery, and, therefore, at that time and place, the aërial electromotive force amounted to that of thirty-eight Daniel's cells."*

Such an approach would have straightforwardly demonstrated the existence of a positive potential gradient[2] in fair weather conditions, but comparable apparatus to replicate this approach - particularly the electrometer - is hard to find in a modern school or undergraduate laboratory.

A simple educational instrument to provide stimulating observations of the fair weather electrical atmosphere is the goal here. The approach and principles of the instrument are described[3]. The method used has classical origins in atmospheric electricity, based on an instrument devised by C.T.R.Wilson, described in a paper published about century ago (Wilson, 1906). Wilson's instrument measured the potential gradient (PG) and vertical air-earth current density $J_S$ simultaneously. This approach was so successful that it was used by the Met Office at their Kew Observatory, near London between 1931 and 1979. The site, instrument, method of measurement and observations obtained have been recently described by Harrison and Ingram (2005).

The new educational instrument presented here operates on the same principles as the Wilson instrument, but uses modern sensitive electronics to produce a small and portable instrument which is readily constructed. A feature is that the copper circuit board it employs includes both the printed sensor electrode and the simple signal processing electronics. The instrument measures both the PG and the vertical air-earth current density, and is called the **E**lectric

[2] Converted to modern units (assuming each zinc-copper Daniel cell generates an emf of 1.08V), the Potential Gradient near Aberdeen, Scotland at 8am on the 14th September, 1859 would have been +137V.m$^{-1}$.
[3] The full constructional details are available at http://arxiv.org/abs/physics/0701280

**P**otential gradient **A**nd **C**urrent (*EPAC*) sensor, to highlight the two atmospheric electrical parameters it measures.

## 2. Operating principle

The EPAC instrument operates by determining the potential on an isolated horizontal plate electrode, which is mounted flush with the ground in dry conditions. The potential is measured using a simple semiconductor electrometer (*e.g.* Harrison, 1996), with respect to the potential of the ground. As for Wilson's apparatus, the measurements can be used to determine the PG from the magnitude of the induced voltage, and, from the rate of change of the induced voltage and the electrode plate's area, the air-earth current flowing.

If a horizontal conducting plate electrically isolated from the ground is exposed to an electric field a charge will be induced on its surface. This charge is proportional to the field it is exposed to by the following equation:

$$Q = \varepsilon_0 AF \tag{1},$$

where $Q$ is the induced charge, $\varepsilon_0$ the relative permittivity, $A$ the plate's surface area and $F$ the electric field the plate is exposed to. The change in voltage on the plate resulting from this induced charge is related to the electrical capacitance of the system by the relationship

$$V_{step} = \frac{Q}{C} \tag{2},$$

with $V_{step}$ being the change in voltage of the plate and $C$ is the capacitance. Combining equations (1) and (2) produces the following:

$$V_{step} = \frac{\varepsilon_0 A}{C} F \tag{3}$$

By measuring the change in voltage of a conducting plate as it is exposed to the atmosphere it is therefore possible to calculate the PG using equation (3) once $V_{step}$ has been calibrated experimentally for a known PG or if $A$ and $C$ are determined for the instrument.

For as long as the plate is exposed upwards to the atmosphere, it will collect positive electric charge brought downwards by the air-earth current density $J_S$. This charge will increase the potential on the plate by a voltage $V_{charge}$ defined by the following equation:

$$V_{charge} = \frac{J_S A}{C} \cdot t \tag{4}$$

with $t$ the time for which the plate is exposed to the atmosphere. This equation has been derived from equation (2) with $Q$ substituted by the total volume of charge carried to the plate by current density $J_S$ to the plate's area A, in time $t$.

If the plate is shielded from the atmospheric PG and $J_S$, it will remain at a constant potential if there is no leakage. However, when the shielding is removed, the potential will immediately increase, with the voltage step proportional to the atmospheric PG in accordance with equation (3). The plate potential will then rise slowly under the influence of $J_S$, as well as fluctuating due to changes in PG.

The EPAC sensor provides a sensing plate of fixed geometry and electrode capacitance, and an electrometer to measure the potential on the plate.

## 3. Implementation of the EPAC sensor

The prototype EPAC sensor was constructed using copper-coated fibreglass circuit board, measuring 9cm by 14cm, with copper surfaces on both sides. The lower side is earthed (see figure 1a), and the upper side carries the printed sensing plate electrode. As with the original Wilson design, a guard ring was etched around the plate electrode, and driven at the measured electrode potential to suppress current leakage.

The potential on the upper sensing plate is measured with respect to ground using a sensitive, ultra-high impedance electrometer chip, which draws a negligible bias current from the plate. This electrometer voltmeter serves only to provide a measurement of the plate voltage at high impedance, to drive low impedance measurement instruments with only unit gain. This therefore provides a drive voltage for the guard electrode. An output potential divider is provided so that the typical voltages induced under fair weather atmospheric conditions fall with the range of typical digital voltmeter. No filtering is required before the op-amp, as any high-frequency noise is removed by capacitive smoothing inherent in the system (e.g. the capacitance provided by the sensing electrode being in close proximity to the ground). A schematic of the electrometer voltmeter system is given in figure 1b.

To measure PG using the EPAC sensor, the sensor was placed on flat, open ground as close to the surface as possible (to avoid perturbing the ambient field). An earthed metal cover was placed approximately 30cm above the plate and the output voltage read. Then the shield was removed and the voltage induced in accordance with equation (3) measured immediately upon exposure. This voltage step ($V_{step}$) was proportional to the PG. After a short (of order one minute) time of plate exposure, the shield was replaced over the sensor and the output voltage measured again. The difference in voltage between the initial reading and the final is $V_{charge}$ and can be used to find $J_S$ from equation (4).

**4. EPAC sensor calibration and comparison**

Calibration of the PG aspect of the EPAC sensor was achieved by measuring the voltage step $V_{step}$ before and after plate exposure, which was compared with corresponding values of PG at 1m made nearby by a calibrated electrostatic field mill (Bennett and Harrison, 2006a). The proportionality is linear and is shown in figure 2.

The equation of the line of best fit was used to convert the voltage step into a value of PG. The close similarity between the field mill (FM) and EPAC derived PG can be seen in figure 3. The instruments were separated horizontally by approximately 10m and 2m vertically during the calibration. The small deviations from linearity between the two instruments seen in figure 2 may therefore be attributed to spatial variation in atmospheric space charge and air of different conductivity carried by turbulent eddies, the effects of which are always evident and are the main cause of observed high-frequency variation in PG during fair-weather.

From the calibration, it is possible to derive the instrument capacitance $C$, which was 235pF. Because of the fixed geometry of the circuit board and printed electrodes, $C$ is expected to be similar for other devices made to the same specification. If $C$ is to be calculated theoretically, the relative permittivity and thickness of the circuit board must be determined (the two plates effectively act as a parallel-plate capacitor, with the circuit board being the medium in between).

The value of $C$ is required for the determination of $J_S$ by equation (4). Using this value for $C$, the one minute mean $J_S$ value was derived using the EPAC sensor, by exposing the plate for one minute. These values are shown as a short time series in figure 4. The mean value for $J_S$ of 1.5pAm$^{-2}$ during the sampling period was compared to $J_S$ measured directly by other instruments (principally the Air-Earth Current Pyramid, detailed by Bennett and Harrison 2006b) nearby, and found to agree closely. Although the EPAC sensing plate is raised above Earth potential due to the arrival of charge during exposure, the increase is small (order 3mV for one minute exposure) compared to atmospheric PG (order 100V/m) so in this respect the sensing plate will not differ significantly from Earth potential, permitting the corresponding $J_S$ measurement to be a close approximation of that arriving at the surface.

The PG and $J_S$ are therefore calculated as follows:

1. Place the EPAC sensor parallel to the ground and cover with an earthed conducting plate to shield the sensor from ambient PG and $J_S$. Record the output voltage.

2. Remove the shield so the sensing electrode is fully exposed to the environment and record the subsequent change in output voltage. This is $V_{step}$ and is used to calculate PG by rearrangement of equation (3).

3. Continue exposure of the sensor for a known duration, then cover again with the earthed shield and note the difference between this voltage and the initial voltage found at stage 1 before exposure. This is $V_{charge}$ and is used to calculate $J_S$ by rearrangement of equation (4).

## 5. Conclusions

The EPAC sensor provides a simple approach to measurement of atmospheric electrical properties. Measurements of PG and $J_S$ using the EPAC sensor agree well with those made using more sophisticated instruments, making this sensor ideal for providing *in-situ* spot recordings of atmospheric electrical measurements.

The EPAC sensor is well suited for use as a tool for educational outreach as it is inexpensive and can be safely operated by school pupils to measure real parameters in environmental electrostatics. It relies on direct, straightforward principles of measurement that can be easily explained, as well as providing insights into electronic construction techniques.

There are many possible applications for the results from the EPAC sensor, such as investigation of the electric field surrounding charged objects in the lab (so no environmental PG present), induced charge on droplets, or perturbation of the ambient PG in the vicinity of grounded objects (*i.e.* trees, buildings etc). If measurements are taken over a sufficiently long time period, the qualitative variations of atmospheric PG due to (1) local aerosol number concentration (*i.e.* increased PG during times of increased local traffic activity or the presence of fog) or (2) the large variation in PG during the overhead passage of convective cloud (which may also result in negative PG being observed) can be seen. The correlation between PG and $J_S$ could be investigated, with the concept of a radiating electrostatic field and "dynamic" fields generated from the flow of charge through a resistor (such as for PG). Furthermore, students may be invited to suggest the origin of the downward-flowing air-earth current density $J_S$, leading to discussion of the concept of the global atmospheric electric circuit.

**Acknowledgements**
AJB acknowledges a studentship from the Natural Environment Research Council (NERC). G. Rogers provided technical support in the instrument development.

**Figures and figure captions**

Figure 1

(a)

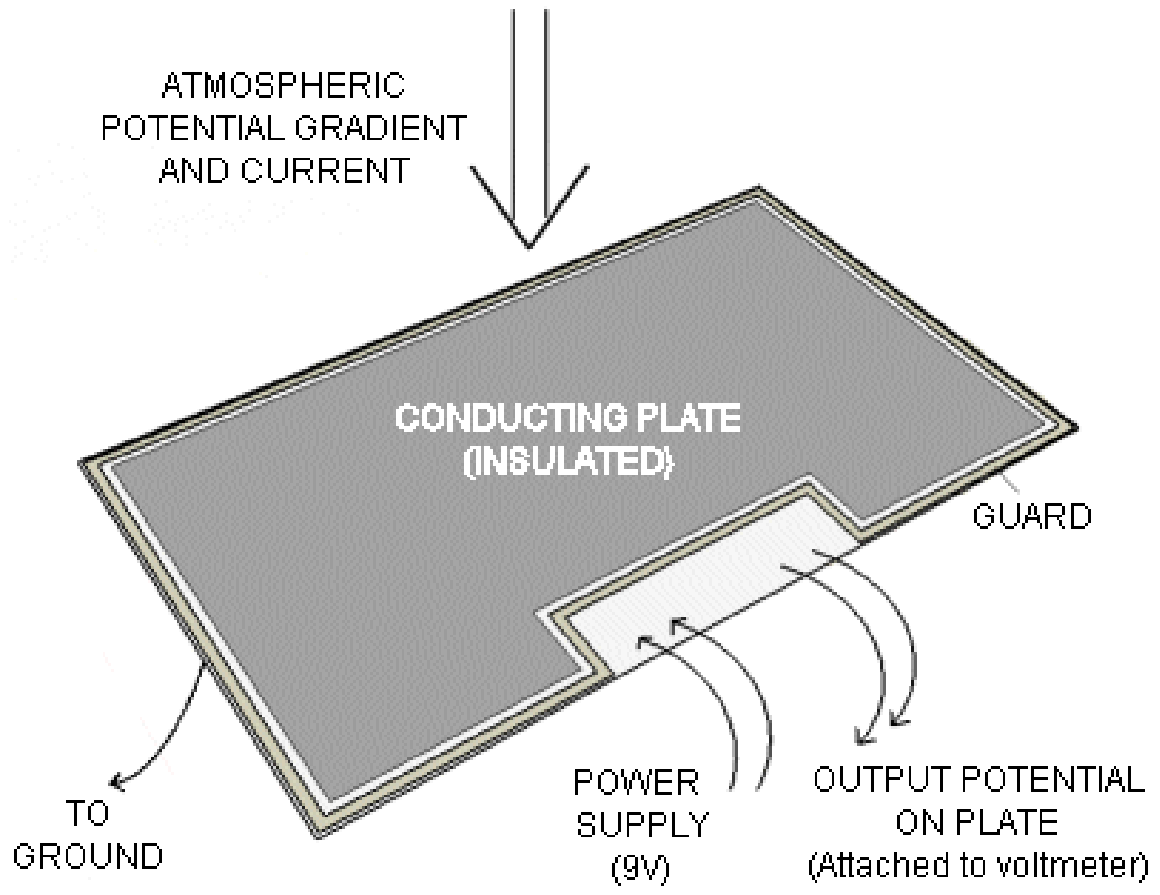

(b)

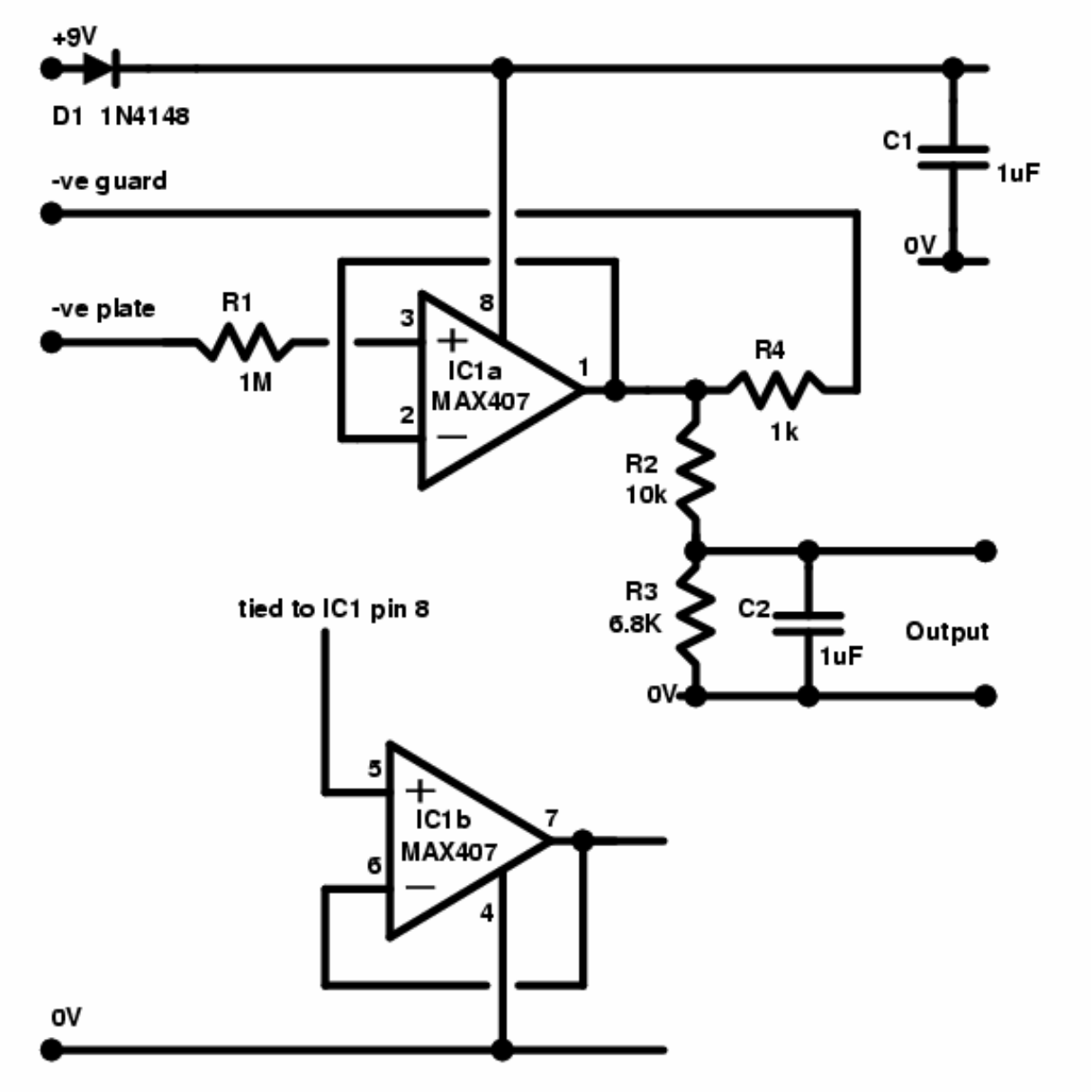

**Figure 1**: (a) Conceptual schematic of the EPAC sensor. The lower face of the sensor is earthed, with the upper face providing a sensing electrode, insulated from earth. The area of the electrode is $8x10^{-3}m^2$. When exposed to the atmospheric potential gradient, a voltage is induced on the sensing electrode, which is proportional to the PG. As long as the electrode is exposed, charge accumulated due to the air-earth current will raise the electrode's potential. An outer guard ring is etched around the sensing electrode, for leakage cancellation. The potential of the plate is determined using a standard specification voltmeter and electrometer circuit. (b) Schematic of the electrometer circuit used with the sensing plate electrode of the EPAC sensor. IC1a is a unit-gain electrometer voltmeter follower, connected to the electrode through protection resistor R1. The low impedance output of the voltmeter follower provides, via protection resistor R4, drive for the electrode's guard ring. R2 and R3 create a potential divider, which reduces the output voltage to a range suitable for a digital voltmeter, with smoothing of high frequency changes provided by C2. Power is provided by a 9V battery, through polarity protection diode D1. Component values are marked on the schematic. (The second opamp in the dual package, IC1b, is not used.)

Figure 2

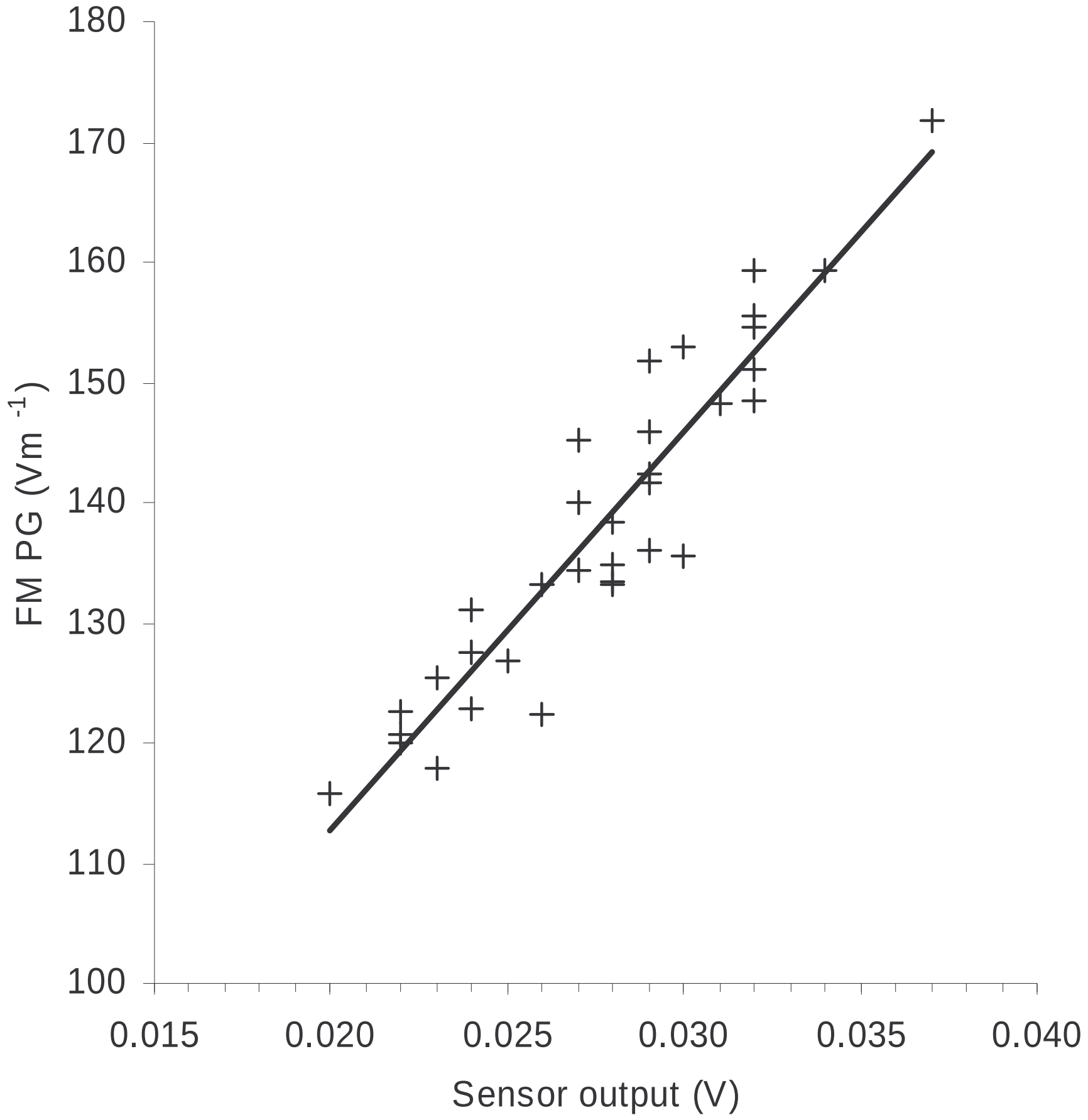

**Figure 2**: Scatter plot of EPAC sensor output $V_{step}$ (measured at the surface) versus the potential gradient (PG) obtained from a nearby field mill (FM) operated at 1m, and sampled at one minute intervals. The correlation coefficient was 0.936.

Figure 3

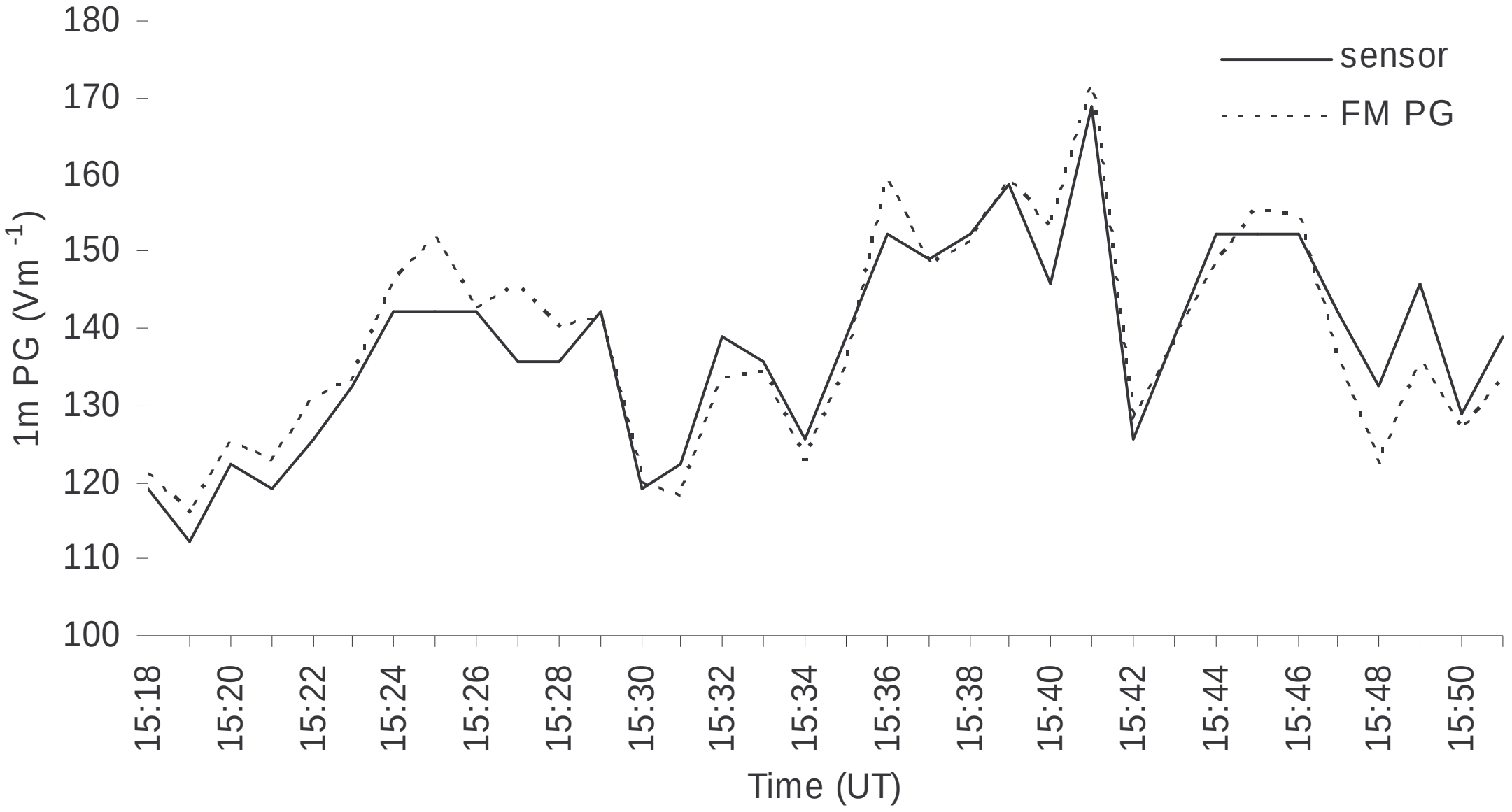

**Figure 3**: Potential gradient (PG) measured at one minute intervals by the field mill (FM) 1m above the surface (dotted) and calibrated EPAC sensor on the surface (solid).

Figure 4

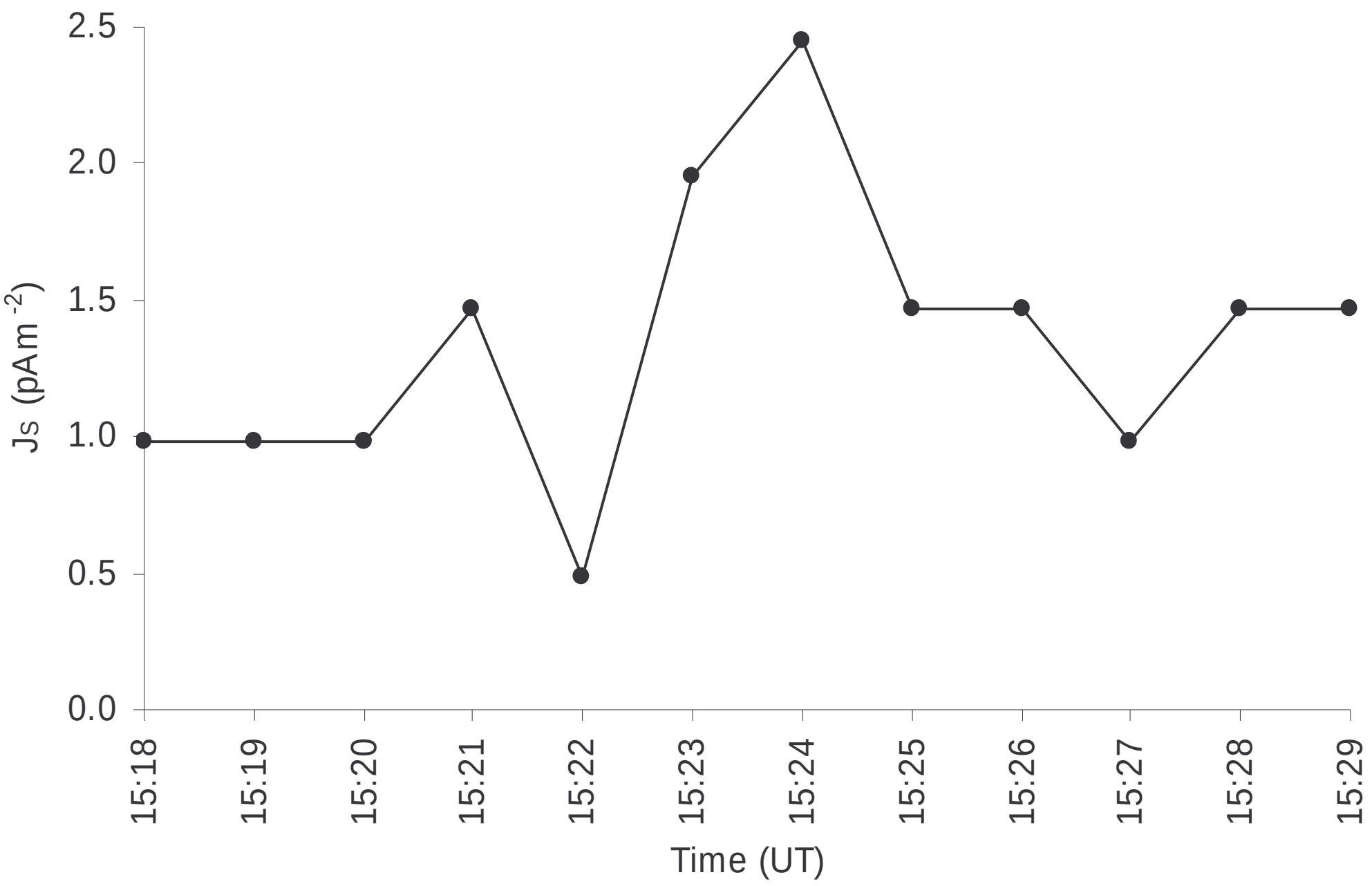

**Figure 4**: One minute values of the mean air-earth current density ($J_S$) measured by the EPAC sensor.